\documentclass{doublecol-new}
\usepackage{natbib,stfloats}
\usepackage{mathrsfs}
\usepackage{graphicx, url, subfigure, longtable, setspace,amsmath, listings, color, enumerate}
\usepackage{natbib}
\usepackage{acronym}

\usepackage{algorithm}
\usepackage{algorithmic}

\theoremstyle{TH}{

}

\theoremstyle{THrm}{

}

\theoremstyle{THhit}{

}

\makeatletter
\def\theequation{\arabic{equation}}

\makeatother

\begin{document}%
%%%%%%%%%%%%%%%%%

\setcounter{page}{1}

\LRH{}

\RRH{Measuring the similarity of PML documents with RFID-based sensors  }

\VOL{x}

\ISSUE{x}

\PUBYEAR{xxxx}

\BottomCatch

%\CLline

\PUBYEAR{2012}

\subtitle{}

\title{Measuring the similarity of PML documents with RFID-based sensors}

\authorA{WANG Zhong-qin, YE Ning }
\affA{Institute of Computer Science, Nanjing University of Post and Telecommunications, Nanjing, 210003, China  \qquad E-mail: cathery163@163.com}

\authorB{Reza Malekian}
\affB{Advanced Sensor Networks Research Group, Department of Electrical, Electronic and Computer Engineering, University of Pretoria, Pretoria, 0002, South Africa \qquad E-mail: reza.malekian@ieee.org}
%

%\authorA{Fusheng Wang\footnote{Work done while working at Siemens Corporate Research.} }
%\affA{Department of Biomedical Informatics, Emory University
%\newline
%36 Eagle Row, Ste 589, Atlanta, GA 30322, USA}
%
%
%
%\authorB{\footnotesize Cristobal Vergara-Niedermayr\footnote{Work done while working at Siemens Corporate Research.}}
%\affB{Oracle \newline
 %New Jersey, USA}
%
%
\authorC{ZHAO Ting-ting, WANG Ru-chuan}
\affC{Institute of Computer Science, Nanjing University of Post and Telecommunications, Nanjing, 210003, China \qquad E-mail: wangrc@njupt.edu.cn}

\begin{abstract}
The Electronic Product Code (EPC) Network is an important part of the Internet of Things.  The Physical Mark-Up Language (PML) is to represent and de-scribe data related to objects in EPC Network. The PML documents of each component to exchange data in EPC Network system are XML documents based on PML Core schema. For managing theses huge amount of PML documents of tags captured by Radio frequency identification (RFID) readers, it is inevitable to develop the high-performance technol-ogy, such as filtering and integrating these tag data. So in this paper, we propose an approach for meas-uring the similarity of PML documents based on Bayesian Network of several sensors. With respect to the features of PML, while measuring the similarity, we firstly reduce the redundancy data except information of EPC. On the basis of this, the Bayesian Network model derived from the structure of the PML documents being compared is constructed.
\end{abstract}

\KEYWORD{Physical Mark-Up Language; Bayesian Network; Electronic Product Code; RFID.}

\REF{to this paper should be made as follows: WANG, Z.Q., YE, N., Malekian, R., ZHAO, T.T., WANG, R.C.(xxxx) `Measuring the similarity of PML documents with RFID-based sensors', {\it Int. Journal. Ad Hoc and Ubiquitous Computing,
}, Vol. x, No. x, pp.xxx\textendash xxx.}

\begin{bio}

{\bf WANG Zhong-qin} received the Bachelor of Electrical Engineering and Automation (Intelligent Building) from Nanjing Jinling Institute of Technology, Jiangsu, China in 2006. He is currently pursuing the Master degree in Computer Software and Theory from Nanjing University of Posts and Telecommunications. His research interests include data clustering, similarity computation, data management and data analysis of network.

{\bf YE Ning} received the BSc in Computer Science from NanJing University, MSc in School of Computer \& Engineering from Southeast University, and Ph.D. in Institute of Computer Science from Nanjing University of Post and Telecommunications, China, where, she is currently a professor. Professor YE Ning is also a senior member of CCF and a member of pervasive computing specialty committee in China. She was a Visiting Researcher in Department of Computer Science, University of Victoria, Canada in 2010. Her research focuses on the security and information processing of network.

{\bf Reza Malekian} is a Senior Lecturer in Department of Electrical, Electronic and Computer Engineering at University of Pretoria. Previously, he was an Assistant Professor of Computer Science and Engineering at University for Information Science and Technology St. Paul the Apostle, Republic of Macedonia and a Postdoctoral Fellow in Faculty of Computing, Universiti Teknoligi Malaysia (Erasmus Partner).
%He received Ph.D. in Computer Science from UTM, M.Eng. (with honor) in Information Technology Engineering from Iran University of Science and Technology, and B.Eng in Computer Engineering from North University.

{\bf ZHAO Ting-ting} received the Bachelor Degree in Digital Media Technology from Tongda College of Nanjing University of Posts and Telecommunications, Jiangsu, China in 2012. She is currently pursuing the Master degree in Computer Software and Theory at Nanjing University of Posts and Telecommunications.
%Her research interests include social network, Internet of Things, and network security.

{\bf WANG Ru-chuan} is a professor and tutor of Ph.D.candidate of Nanjing University of Posts and Telecommunications,interested in VR,graphics and image,the security of network and mobile agent technology,computer network and grid computing and sensor network.

\end{bio}

\maketitle

 \section{Introduction}

The Internet of Things (IoT) is a novel network architecture that is rapidly gaining attention in the scenario \cite{1}.The Physical Mark-Up Language (PML) is a collection of common, standardized XML vocabularies to represent and describe information related to EPC Network enabled objects as presented by \cite{2}. The documents of each component to exchange data in EPC Network system are XML documents based on PML Core schema, and this type document is called the PML document as described by \cite{3}. Among them, the purpose of the PML Core schema is to provide a standardized format for the exchange of the data captured by the sensors in an Auto-ID infrastructure, e.g. RFID readers. PML is to be regarded as the complementary vocabularies for business transactions or any other XML application libraries, which include a new library composed of relevant definitions about EPC Network system, rather than to replace the XML to be a new markup language.

EPC tags to identify each of objects are adapted by Auto-ID Center. Different sorts of sensors which equipped on shops, warehouses, workshops and so on \cite{4}, are to acquire EPC data and other information, such as temperature and geography location. It is essential for EPC network to process these data signed by PML documents at speed of hundreds of millions per second. For managing theses huge data stream and reduce network traffic, it is inevitable to develop the high-performance technology for managing these PML documents, such as implementing a joint forensics-scheduling scheme\cite{5}, applying a multi-layered algorithm to manage real-time data \cite{6}, or compressing the amount of data, filtering and integrating these tag data.

For above purpose, one of the effective methods is clustering as presented by \cite{7}, which could depend on the structure and semantics of these data. Indeed, the similarity computation, which measures the similarity of the compared PML documents, is the foundation of the clustering method. In this paper, we mainly focus on the similarity computation of PML documents.
Many researches have proposed a wide range of algorithms for XML similarity computation, the kind of technique being used mainly include ED-based (Tree Edit Distance), IR-based (Information Retrieval) and others (e.g., edge matching, path similarity, etc.) to measure similarity of the XML documents.

Some of above methods of XML similarity mainly concern on the structural properties of XML data and disregard element/attribute values of XML\cite{8}, but many others consider values in their similarity computations. With respect to XML documents which are less structurally disparate (they might originate from the same data source, and might even conform to the same grammar), similarity computation based on structure and content is a favorable method \cite{9}. As follow, we introduce algorithms of structure-and-content method.

Liang and Yokota\cite{10} provided an approximate XML similarity method based on leaf nodes (leaf node values in particular), entitled LAX (Leaf­clustering based Approximate XML join algorithm). Kade and Heuser\cite{11} developed a method for comparing XML documents as documents lists. Weis and Naumann\cite{12}put forward a method entitled Dogmatix for comparing XML elements (and consequently documents) based on their direct values, as well as corresponding parent and children similarities. An approach for document/pattern comparison, developed in the context of data integration and XML querying, is proposed by Dorneles et al.\cite{13}. Leitao\cite{14} provided a probabilistic approach, using a Bayesian network to combine the probabilities of children and descendents being duplicates, for a given pair of XML elements in the documents being compared. The similarity between two XML documents corresponds to the probabilities of their root nodes being duplicates.

PML document management in highly dynamic environments in EPC network systems is a hard task. In this environment, documents change very frequently, both in content and structure. Additionally, the same information may be represented in documents from different sources, leading to (partial or total) overlap of documents. Dealing with these overlaps and/or duplications in a dynamic environment is challenging in many aspects. In this paper, we improve the method of XML Fuzzy Duplicate Detection proposed by Leitao in accordance with the features of PML document, and propose a method of measuring the similarity of PML documents based on Bayesian Network. It not only considers the duplicate status of children, but rather the probability of descendants being duplicates. With respect to the features of PML, while measuring the similarity, we firstly reduce the redundancy data except information of EPC. On the basis of this, the Bayesian Network model derived from the structure of the PML documents being compared is constructed. And this model has taken into consideration not only the EPC values contained in the PML but also their internal structure. Then the similarity between two PML documents could be deduced. Finally, simulations show the proposed algorithm is able to maintain high precision and recall values.

The remainder of the article is organized as follows. In Sec.2, we describe the background of The remainder of the article is organized as follows. In Sec.2, we describe the background of PML documents and Bayesian network. In Sec.3, we present the Bayesian network for PML similarity computation, including the relationship between PML documents similarity and Bayesian network probability, redundancy reduction of PML documents, Bayesian network model for PML documents and the algorithm of constructing Bayesian network model, and elucidate how PML similarity measure is performed using the proposed Bayesian network. Section 4 presents our prototype and simulative tests. Section 5 concludes the paper and outlines future research directions.

\section{Background}
\subsection{PML document}
In order to stress the need for relatedness assessment in PML document comparisons, we consider the example in Fig. 1(a) and Fig. 2(a). It depicts PML document of data captured by RFID readers. RFID readers capture the Electronic Product Code stored on the individual Auto-ID compliant tags (e.g. 1:2.24.404 and 1:12.8.128).

PML document should enable to elaborate the process that RFID readers acquire data,including where is the certain RFID reader, which is identified by a unique identifier (e.g. 1:4.16.36), when certain tags in its read range are observed (e.g. 2002-11-06T13:04:34-06:00) and so on. Each such observation might need to be labeled with the command that was issued to trigger the observation (e.g. READ\_PALLET\_TAGS\_ONLY) and a unique label to reference a certain observation (e.g. $00000001$).

Within the EPC Network, RFID readers are one of the main components. The data they capture are routed within the EPC Network from readers to Savant as described by \cite{15} (the Savant is a middleware system which requests from upper application and receives data from sensors.) ,from one Savant to other, from Savant to the EPC Information Service. To standardize the mark-up of those captured data, PML document needs to adequately represent the observed values.

XML documents represent hierarchically structured information and can be modeled as Ordered Labeled Trees (OLTs)\cite{16}. In the OLTs, nodes represent XML elements and are labeled with corresponding element tag names. Element attributes mark the nodes of their containing elements. Some studies have considered OLTs with distinct attribute nodes, labeled with corresponding attribute names\cite{17}. Attribute nodes appear as children of their encompassing element nodes, sorted by attribute name, and appearing before all sub-element siblings\cite{18}. So we reference the XML document’s OLTs and describe the PML document’s OLTs in Fig.1 (b) and Fig.2 (b). Element/attribute values are also considered in the comparison process following the application of structure­and­content. As an example, consider the tree representation of two PML elements represented in Fig. 1(b) and Fig. 2(b) (Nodes are labeled by their PML tag name and an index for future reference). Both trees represent XML elements named sensor. They nest further XML elements representing ID and Observation. An Observation consists of ID, Command, DateTime and Tag, represented as children PML elements of Observation. And a Tag also consists of ID and other elements which might involve other children elements. All of those elements have a text node which stores the actual data. For instance, DateTime has a text node containing ‘2002-11-06T 13:04:34-06:00’ as string value.

\begin{figure}
\centering
\includegraphics[width=8cm, height=12cm]{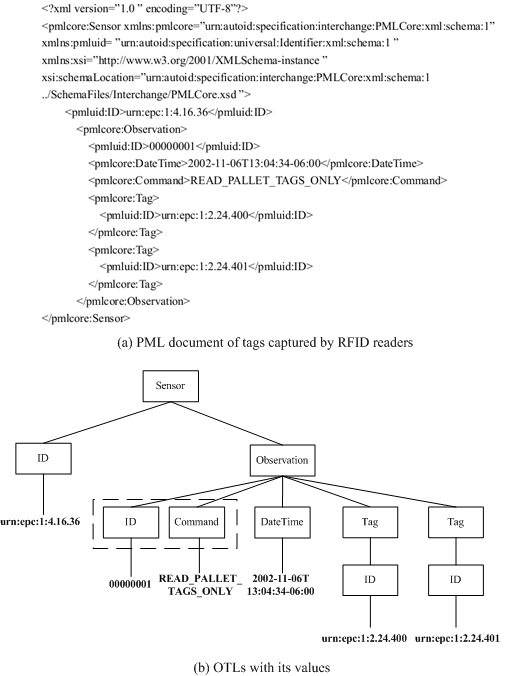}
\caption{Representation of PML document of tags captured by RFID readers }
\end{figure}

\begin{figure}
\centering
\includegraphics[width=8cm, height=12cm]{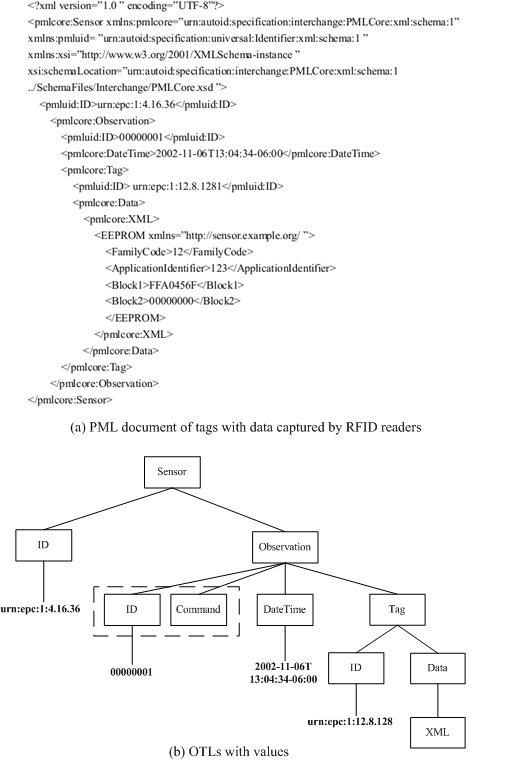}
\caption{Representation of PML document of tags with data captured by RFID readers }
\label{Representation of PML document of tags with data captured by RFID readers}
\end{figure}

\subsection{Bayesian network }
Bayesian networks (BNs) provide a graphical formalism to explicitly represent the dependencies among the variables of a domain, thus providing a concise specification of a joint probability distribution as described by \cite{16, 17}. The network structure of the Bayesian network (belief networks or Bayes nets for short), belonging to the family of probabilistic graphical models (GMs), is an DAG (Directed Acyclic Graph), where each node represents an attribute or data variables and the arcs represent the probabilistic dependency relation between attribute nodes. The relationship of complex variables in specific issues is represented by a network structure, reflecting dependency relationship between variables in the problem areas. In addition to the DAG structure, which is often considered as the "qualitative" part of the model, one needs to specify the "quantitative" parameters of the model. The parameters are described in a manner which is consistent with a Markovian property, where the conditional probability distribution (CPD) at each node depends only on its parents as presented by \cite{18}. A mathematic model is used to express Bayesian network as follows:
\begin{equation}
B=(V, E, P)
\end{equation}
The set of collection of random variables is defined as:
\begin{equation}
V=(V_{1}, V_{2},...,V_{n})
\end{equation}
The collection of directed edges is defined as:
\begin{equation}
E=(V_{i}V_{j}| V_{i},V_{j} \in V)
\end{equation}
The set of Conditional probability distribution, namely Conditional probability table is defined as:
\begin{equation}
P={P(V_{i}|V_{1},V_{2},..., V_{i-1}, V_{i} \in V)}
\end{equation}
Consider the following example that illustrates some of the characteristics of BNs. The example shown in Figure 3 presents the Bayesian network of two PML documents being the rooted node of sensor, which have the same data structure but different value. Firstly, it considers Tag similarity, represented by the variable Tag (denoted by ST) might result from $ID^{'}$ similarity, represented by the variable $ID^{'}$ (denoted by $SI^{'}$). Secondly, Observation similarity represented by the variable Observation (denoted by SO) might result from DateTime similarity represented by the variable DateTime (denoted by SD). In the final case, it is reasonable to assume that sensor similarity represented by the variable Sensor (denoted by SS) will be determined by SO and ID similarity, represented by the variable ID (denoted by SI). All variables are binary; thus, they are either true (denoted by "T") or false (denoted by "F").
\begin{figure}[h]
\centering
\includegraphics[width=9cm, height=6cm]{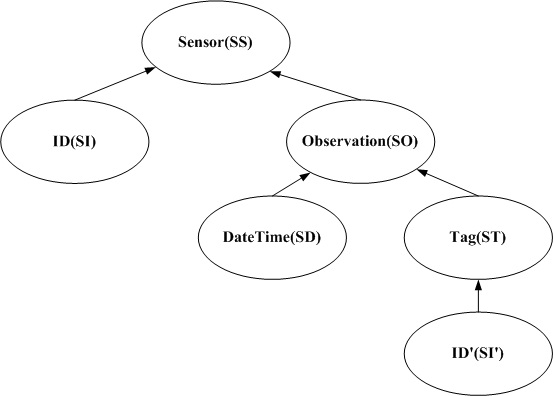}
\caption{Bayesian network of two sensors}
\label{Bayesian network of two sensors}
\end{figure}
For example, the CPTs of Tag and ID' are listed in Table 1 and Table 2. P(SI') presents the same probability of two ID's and P(ST) presents the same probability of two Tags.
\begin{table}[htbp]
	\caption{ Same probability of two ID's}
	\label{table_five}
	\begin{center}
	\begin{tabular}{|c|c|} \hline
	%& $Interface Name$ & $Link Local Address$    \\ \hline
	P(SI'=F) & P(SI'=T)   \\ \hline
	0.5 & 0.5   \\ \hline
	
	\end{tabular}
	\end{center}
\end{table}

\begin{table}[htbp]
	\caption{ Same probability of two Tags with the same probability of two ID's}
	\label{table_five}
	\begin{center}
	\begin{tabular}{|c|c|c|} \hline
	%& $Interface Name$ & $Link Local Address$    \\ \hline
	SI' & P(ST=F) & P(ST=T)   \\ \hline
	F & 1 & 0   \\ \hline
	T & 0 & 1   \\ \hline
	
	\end{tabular}
	\end{center}
\end{table}
From total probability formula,
\begin{equation}
P(B)=\sum_{i=1}^{n} P(A_{i})P(B|A_{i})
\end{equation}
We enable to demonstrate that the different probability of two tags P(ST=T) is 0.5, while the probability is defined as:
\begin{eqnarray}
P(ST=T)=P(ST=T|ST'=T)P(SI'=T')\\
\nonumber +P(ST=T|SI'=F')=1 \times 0.5 + 0\times 0.5=0.5
\end{eqnarray}
This results in:
\begin{equation}
P(ST=F)=1-P(ST=T)
\end{equation}
Similarly, by applying Eq. (1), probability $P(SS=T)$ is defined as:
{\footnotesize
\begin{eqnarray}
\footnotesize P(SS=T)=P(SS=T|SI=T)P(SI=T)+\\
\nonumber P(SS=T|SI=F)P(SI=F)+P(SS=T|SO=T)\\
\nonumber P(SO=T)+P(SS=T|SO=F)P(SO=F)
\end{eqnarray}}
And is result,
\begin{equation}
P(SS=F)=1-P(SS=T)
\end{equation}

\section{Bayesian network for PML similarity computation}
\subsection{Redundancy reduction of PML documents}
The aim of this phase is to reduce the redundant nodes in the original tree before construction of Bayesian network.
After researching the PML Core specification defined in 'PMLCore.xsd' XML schema file, we know that the rooted element Sensor is main comprised of two subordinate ID element and Observation element. And the Observation element consists of the following:
\begin{itemize}
\item an optional ID element
\item an optional Command element
\item DateTime element
\item zero or more Data elements
\item zero or more Tag elements
\end{itemize}
Among them, the Tag element consists of the following elements:
\begin{itemize}
\item ID element
\item optional Data element
\item zero or more Sensor elements
\end{itemize}

A sensor is considered as any devices that make measurements and observations,such as an RFID reader or a temperature sensor. As mentioned earlier, each of objects, including different sensors, has a unique ID, namely EPC, to identify their information in EPC network. EPC regarded as a point enable to inquiry and retrieve information from supply chains. In the paper, we mainly concern on the similarity of PML documents rather than the concrete information that each PML document contain. For example in Fig. 2(a), the information stored in tag EEPROM is not important for PML comparison similarity. If the client wants to acquire these data, they enable to receive the EPC by RFID reader, which finding IP address to get the object information stored in EPC IS from internet. So redundancy is the data in addition to be able to identify EPC, including ID, Command, DateTime and Data in Observation element and Data , Sensor in Tag element, only retaining the ID in Tag element.
Redundancy reduction of tree deletion operations between two rooted ordered labeled trees that represent two PML documents are defined as follows:
\begin{itemize}
\item Given a leaf node x and a tree $T$, $T$ containing node $p$ with first level sub-trees and $x$ being the $i$th child of $p$, e.g. ${P_{1},...P_{i-1},x,P_{i+1},...,P_{m}}$ $DelLeaf(x, p)$  is the deletion operation applied to node p that yields x with first level sub-trees ${P_{1},...P_{i-1},P_{i+1},...,P_{m}}$ (Figure 4).
\item Given a sub-tree $A$ and a tree $T$, $T$ containing node $p$ with first level sub-trees, e.g.${P_{1},...P_{i-1},A,P_{i+1},...,P_{m}}$ $DelLeaf(A, p)$ is the deletion operation applied to node $p$ that deletes sub-tree $A$ in $T$ from among the children of ${P_{1},...P_{i-1},P_{i+1},...,P_{m}}$ (Figure 4).
\end{itemize}

\begin{figure}
\centering
\includegraphics[width=9cm, height=6cm]{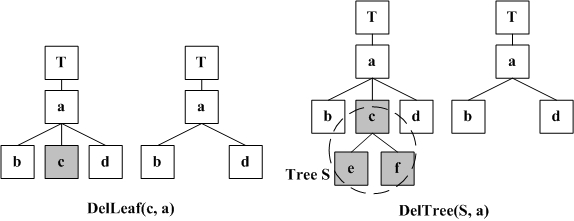}
\caption{Delete leaf node and Delete sub-tree}
\label{Delete leaf node and Delete sub-tree}
\end{figure}
In our model, we first simplify a PML tree using the algorithm Pred. The description of algorithm is as follows.
\begin{algorithm}
\caption{Pred(PMLtree T)}
01:foreach node Ni in NodeList do \\
02:if Ni==Observation then \\
03:     foreach childnode t of Observation do \\
04:if t {ID, Command, DataTime} then \\
05:             DelLeaf(Observation, t); \\
06:         else if t==Tag then \\
07:               foreach childnode s of Tag do \\
08:if s==Data then \\
09:                       DelTree(Tag, Data);

\end{algorithm}
The input of the algorithm is a PML tree, as shown in in Fig.1 (b) and Fig. 2 (b). We assume that all nodes are stored in a dynamic list NodeList in accordance with the gradation in the tree. And the parent-child relationship between the nodes is also shown in the list. The algorithm traverses the list NodeList and at the same time, using the functions of DelLeaf and DelTree to respectively delete the redundant nodes and subtrees. Definition of DelLeaf $(c, a)$ is to delete a leaf node c that is eligible for deleting and is parented at node a. What's more, $DelTree (S, a)$ is used to delete a eligible subtree S that is parented at a.

The result of the algorithm is to obtain a new NodeList made of the remaining nodes by the way of deleting those redundant nodes and subtrees. Of course, the deleting operation will not change the original gradation relationship. The output is shown in Figure 5.
\begin{figure}
\centering
\includegraphics[width=9cm, height=6cm]{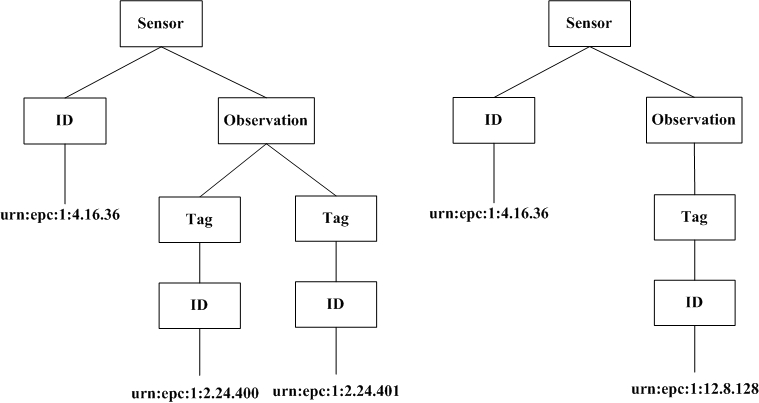}
\caption{Result of redundancy reduction from PML documents in Fig.1 (a) and Fig. 2 (a)}
\label{Result of redundancy reduction from PML documents in Fig.1 (a) and Fig. 2 (a)}
\end{figure}

\begin{figure}
\centering
\includegraphics[width=7cm, height=7cm]{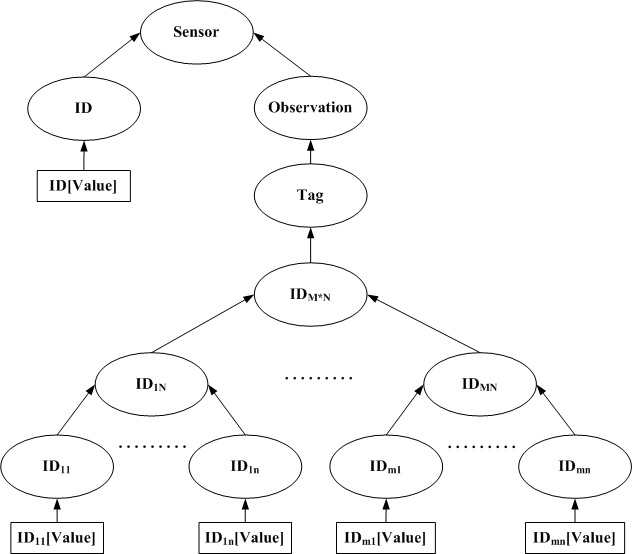}
\caption{Bayesian network model}
\label{Bayesian network model}
\end{figure}
\subsection{Bayesian network model for PML documents}
For measuring the similarity of PML documents, we construct a Bayesian network model as illustrated in Fig. 6. The network model has a rooted node labeled Sensor representing the possibility of node sensor in two compared PML trees. If a tree A and tree B are two compared trees, the node Sensor represents the possibility of node Sensor in tree A being a duplicate of the node Sensor in tree B. The probability of the Sensor nodes being duplicates depends on the probability of each pair of children nodes being duplicates. Then the node ID represents the possibility of node ID in tree A being a duplicate of the node ID in tree B; node Observation represents the possibility of node Observation in tree A being a duplicate of node Observation in tree B. Similarly, we enable to repeat the process of other two nodes.

However, it is a slightly different procedure of PML nodes labeled ID of the children of Tag node. In this case, we wish to compare the full set of nodes, instead of each node independently. In this case, the set of ID nodes of the children of Tag nodes being duplicate depends on each ID node in tree $A$ being a duplicate of any ID node in tree $B$. It is presented by nodes $ID_{M*N}$, $ID_{MN}$ and $ID_{in}$ in Fig.6. Because the nodes $ID_{in}$ have no children, their probability of being duplicates only depends on their values $ID_{in}[Value]$.

We know that elements of Sensor, ID and Observation are contained in each of PML documents from the PML Core schema. And the probability of the two PML nodes being duplicates depends on (1) whether or not their values of nodes are duplicates, and (2) whether or not their children of nodes are duplicates. The node is assigned a binary random variable. If a node exists in the same location of two PML trees, this variable takes the value 1 to present. Otherwise, the variable takes the value 0 to express.
With respect to the Bayesian network model, we could compute the probability in Fig. 1 (a) and Fig. 2 (a). Three types of conditional probabilities are defined as follows:
\begin{itemize}
\item The probability of the values of the nodes being duplicates depends on each individual pair of values being duplicates;
\item The probability of two nodes being duplicates depends on their values and their children being duplicates or each pair of children nodes being duplicates (i.e. Sensor).
\item The probability of a set of nodes of the same type being duplicates depends on each pair of individual nodes in the set are duplicates.
In our example, these two types of conditional probabilities correspond to the respective probabilities listed in Tables 3, 4, 5.
\end{itemize}

\begin{table}[htbp]
	\caption{ Conditional Probabilities}
	\label{table_five}
	\begin{center}
	\begin{tabular}{|c|c|c|} \hline
	%& $Interface Name$ & $Link Local Address$    \\ \hline
	Conditional Probability   \\ \hline
	$P(ID| ID [Value])$   \\ \hline
	$P(ID_{mm}| ID_{mm} [Value])$   \\ \hline
	\end{tabular}
	\end{center}
\end{table}

\begin{table}[htbp]
	\caption{ Conditional Probabilities}
	\label{table_five}
	\begin{center}
	\begin{tabular}{|c|c|c|} \hline
	%& $Interface Name$ & $Link Local Address$    \\ \hline
	Conditional Probability   \\ \hline
	$P(Tag| ID_{M*N})$   \\ \hline
	$P(Observation| Tag)$   \\ \hline
    $P(Sensor| ID, Observation)$   \\ \hline
	\end{tabular}
	\end{center}
\end{table}

\begin{table}[htbp]
	\caption{ Conditional Probabilities}
	\label{table_five}
	\begin{center}
	\begin{tabular}{|c|c|c|} \hline
	%& $Interface Name$ & $Link Local Address$    \\ \hline
	Conditional Probability   \\ \hline
	$P(ID_{iN}| ID_{i1,..., IDin})$  \\ \hline
	$P(ID_{M*N}| ID_{1N},..., ID_{MN})$ \\ \hline
    	\end{tabular}
	\end{center}
\end{table}

\subsection{The algorithm of constructing Bayesian network model}
In this paper, a PML tree is defined as a triple $T=(S, V, W)$, where
\begin{itemize}
\item  	$S$ is a root node label, e.g., for tree $T$ in Fig.5, $S=Sensor$.
\item  	$V$ node label with (attribute, v) pair, where v is the value of this node. If the node itself has a value, we define it as a special (attribute, v) pair. For tree T in Fig.5, we have a node V with (ID, urn: epc: 1:4.16.36) pair.
\item 	$W$ is a set of PML trees, means that $W$ is the set of subtrees of $T$. These subtrees are again each described as a triple. For tree $T$ in Fig.5 , $W$ contains subtree rooted at observation.
\end{itemize}

\begin{algorithm}
\caption{Merg(PTree T,PTree T')}
01:Input: $T=(S, V, W)$  \\
02:     $T'=(S^{'}, V^{'}, W^{'})$  \\
03:Output: A directed graph $G=(N,E)$ \\
/* -------------- Initialization --------------- */ \\
04:$X= Y =0;$ \\
/* --------------------------------------------- */ \\
05:if $S==S^{'}$ then  // Two root nodes are the same.\\
06:Insert a node $S$ into $N$; \\
07:if $V \cup V' \neq \emptyset$ then   // At least one of the two nodes is not NULL.\\
08:      if $V == V^{'}$ then  // Two nodes are the same including attributes and values.\\
09:Insert a node $V$ into $N$; \\
10:Insert an edge into E from this node to $S$; \\
11:              Insert a node $v$ into $N$; \quad \quad $v$ represents value. \\
12:              Insert an edge into $E$ from this node to $V$; \\
13:              Insert a node $v^{'}$ into $N$; \\
14:             Insert an edge into E from this node to $V$; \\
15:if $W \cup W^{'} \neq \emptyset$ then // At least one of the two sets is not NULL.\\
16:foreach $W_{i} \in W$ do \\
17:foreach $W_{j}^{'}\in W'$ do  //These two nested loops are used to implement one-to-one comparison of all nodes in two sets.\\
18:if $Wi \neq Tag$ and $W_{j}^{'} \neq Tag$ then  //None of them owns a Tag. \\
19:  R=S; \\
20:$G'=(N^{'},E^{'})\leftarrow Merg(Wi,W_{j}^{'})$  // It recursively invoke function Merg.\\
21:foreach node $n \in N^{'}$ do  //The following three loops are used to link the directed graph G'=（N'，E'）of subtree with the already generated G.  \\
22:    Insert n into $N$; \\
23:foreach edge $e \in E'$ do \\
24:   Insert e into $E$; \\
25:foreach node $n \in N^{'}$ without outgoing edges do \\
26:    Insert an edge into $E$ from this node to $R$; \\
27:            else    \quad \quad  (Any of them owns at least a Tag). \\
28:Insert a node Tag into $N$; \\
29:               Insert an edge into $E$ from this node to $S$; \\
30:if $W_{i}$==Tag then   // Count the number of Tag in W.\\
31:                 $X++$; \\
32:               if $W_{j}^{'}$==Tag then  // Count the number of Tag in W'.\\
33:                 $Y++$; \\
34:        Insert a node $ID_{X*Y}$ into ${N}$; \\
350:        Insert an edge into $E$ from this node to Tag; \\
36:foreach Tag $t_{i}(1 \leq i \leq X) \in W$ do  //The following two loops are used to generate about X*Y nodes IDi*j and other X*Y nodes named with the value of each Tag.\\
37:    $P=ID$ value of Tag $t_{i}$; \\
38:              Insert a node $ID_{i*Y}$ into $N$; \\
39:              Insert an edge into $E$ from this node to $ID_{X*Y}$; \\
40:foreach Tag $t_{j}(1\leq j \leq Y)$ $W^{'}$ do \\
41:    Q=ID value of Tag $t_{j}$; \\
42:                  Insert a node $ID_{i*j}$ into $N$; \\
43:                   Insert an edge into $E$ from this node to $ID_{i*Y}$; \\
44:                   Insert a node $P$ into $N$; \\
45:                   Insert an edge into $E$ from this node to $ID_{i*j}$; \\
46:                   Insert a node Q into $N$; \\
47:Insert an edge into $E$ from this node to $ID_{i*j}$;
\end{algorithm}
The idea of designing Algorithm is to merge two PML trees into one tree, which is starting from the root nodes. We assume that two trees can only be merged in the case of that the root nodes are the same. In our example, the root nodes are the identical S=Sensor, while $V$ has only one element (ID, value). And $W$ is the subtree rooted at Observation. There are two variables, $X$ and $Y$, which are respectively used to store the number of Tags in the two trees. It is clear that they are initialized as Null.

In this algorithm, we define the structure of the input PML tree, as described above, a triple. The algorithm takes as input two sets of PML trees $T$ and $T^{'}$(line$01-02$). We only deal with the case of root node $S=S^{'}$(line $05-47$), otherwise we will exit the algorithm and the output is Null. In the former case, we judge $V = V^{'}$ whether to set up. Under the condition of $V = V^{'}$, we respectively construct new nodes named as the values $v$ of elements of $V$ and $V^{'}$(line $11-14$). After that the function will construct a new edge pointed to the root node (line $09-10$). In the subtree, we recursively invoke the merging function Merg (line $15-20$). In the case of meeting with Tag, it is inevitable to make a one-to-one comparison for which requires a new node $ID_{i*j}$ (line $36-47$). So it is necessary to generate nodes with the number of $X*Y$(line $30-35$). The result of this algorithm is a directed graph $G=(N,E)$, where $N$ is the set of nodes in G while E represents the set of edges between these nodes(line $03$). This graph is initialized as NULL (line$04$). When applying this algorithm to the PML tree $T$ and $T^{'}$ of Fig.5, we can obtain the directed graph in Fig.6.

\subsection{Defining the probabilities}
As illustrated in previous section, we describe how to construct the Bayesian network model, so we need to define the conditional probabilities to inner nodes and prior probabilities to leaf nodes. Here we also define the notion $P(x)$ to mean $P(x=1)$, presenting the probability of two same nodes occurring at the same time.
\subsubsection{Conditional probabilities}
{\bf Conditional Probability CP1}: CP1 denotes that the probability of the values of the nodes being duplicates depends on each individual pair of values being duplicates. In this case, we enable to define $P$ $(ID_{t_{ij}} | t_{ij}[n_{1}], t_{ij}[n_{2}],...)$ to correspond to above presentation, where $ID_{t_{ij}}$ is a leaf node ID of parent node $t_{ij}, t_{ij}[n]$ is the value of attribute n of the $i$-th node with tree $t$ in the PML tree.

If all values of attribute $n$ are duplicates, we consider the value of leaf node ID of parent node $t_{ij}$ as duplicates, and this value represents the importance of the corresponding attribute in determining whether the nodes are duplicates. For instance, if the attribute $ID_{11}[Value]$ is equal to 1, then we consider the leaf node $ID_{11}$ values are duplicates.

This definition is represented in Eq. (10), and we determine that the probability of the PML nodes being duplicates equals a given value, $w$.

\begin{equation}
P(ID_{t_{ij}}|t_{ij}[n_{1}],t_{ij}[n_{2}],...)=\sum_{1 \leq k \leq n, t_{ij}[a_{k}]=1} W_{a_{k}}
\end{equation}
Subject to $\sum_{1 \leq k \leq n} W_{a_{k}}$
In this case, since all of leaf nodes have only an attribute value, Equation (11) is represented as follows:
\begin{equation}
P(ID_{t_{ij}}|t_{ij}[n_{1}])=\sum_{1 \leq k \leq n, t_{ij}[a_{k}]=1} W_{a_{k}}=1
\end{equation}
For instance, $P(ID|ID[Value])=1$ and $P(ID_{11}|ID_{11}[Value])=1$

{\bf Conditional Probability CP2}: CP2 denotes that the probability of two nodes being duplicates depends on their values and their children being duplicates or each pair of children nodes being duplicates. i.e., $P(Sensor| ID_{Sensor}, Ob_{Sensor})$, if both ID and Observation values and their children are duplicates, we could consider the nodes as duplicates. So this definition is represented in Eq. (12).
\begin{eqnarray}
P(t_{ij}|ID_{t_{ij}},Ob_{t_{ij}})= = \left\{
  \begin{array}{l l}
    1 & \quad \textrm{if $ID_{t_{ij}}=Ob_{t_{ij}}=1$}\\
    0 & \quad \textrm{Otherwise}
  \end{array} \right.
\end{eqnarray}

{\bf Conditional Probability CP3}: CP3 denotes that the probability of a set of nodes of the same type being duplicates depends on each pair of individual nodes in the set are duplicates, i.e., $P(ID_{M*N}|ID_{1N}, ID_{2N},...)$ and $P(ID_{1N}|ID_{11}, ID_{12},...)$, the set of nodes ID depends on that each of its nodes is a duplicate. We also assume that the more nodes ID are duplicates, the higher the probability that the whole set of nodes is a duplicate. So this definition is represented in Eq. (13).
\begin{equation}
P(t_{M*N}|t_{1*N}, t_{2*N},...)=\frac{1}{n}\sum{k=1}^{n} t_{kN}
\end{equation}

And the probability $P(ID_{1N}|ID_{11}, ID_{12},... )$, which reflects the fact that a node ID in an PML tree is a duplicate if it is a duplicate of at least one node of the same type in the other PML tree. This is represented in Eq. (14).
\begin{eqnarray}
P(t_{iN}|t_{i1},t_{i2},...t_{iN})= = \left\{
  \begin{array}{l l}
    1 & \quad \textrm{if $\exists j |t_{ij}=1$}\\
    0 & \quad \textrm{Otherwise}
  \end{array} \right.
\end{eqnarray}

\subsubsection{Prior probabilities}
Note that the $P(t_{ij}[n])$ can be defined based on the similarity between values, the greater the probability is, the greate the similarity will be. For instance, the probability of the ID attributes in two Sensor  elements being the same can be similar between both ID nodes. We normalize this similarity to a value between $0$ and $1$. Thus, we define
{\tiny
\begin{equation}
P(t_{ij}[n])= = \left\{
  \begin{array}{l l}
    sim(ID_{i}[n],ID_{j}[n]) & \quad \textrm{if similarity
was measured}\\
    1-sim(ID_{i}[n],ID_{j}[n]) & \quad \textrm{Otherwise}
  \end{array} \right.
\end{equation}}
Where $sim( )$ is a similarity function, normalized to fit between $0$ and $1$.
For instance, for the ID attribute in the Sensor nodes, we can define $sim(ID, ID^{'})=1$ if $ID[Value]= ID^{'}[Value]$, and otherwise $sim(ID, ID')=0$.

\subsubsection{Finally probability}
All conditional and prior probabilities are defined, so we could depend on the knowledge of Bayesian network to compute the probability of two PML trees. And the Bayesian network model has been described in sec. 3.2. According to the network, and applying Eq. (12), the probability is defined as:
{\scriptsize
\begin{eqnarray}
P(Sensor)=\sum{ID,Ob} P(Sensor|ID_{Sensor},Ob_{Sensor})\\
\nonumber P(ID_{Sensor},Ob_{Sensor})=\sum{ID, Ob}P(Sensor|ID_{Sensor},\\
\nonumber Ob_{Sesnor})P(ID_{Sensor})P(Ob_{Sensor}) P(ID_{Sesnor}) P(Ob_{Sesnor})
\end{eqnarray}}

Similarly, by applying Eq. (10), probability $P(ID_{Sensor})$ is defined as:
{\small
\begin{eqnarray}
P(Sensor)=P(Sensor|ID_{Sensor}[Value])P(ID_{Sensor}\\
\nonumber [Value])= w_{value}P(ID_{Sensor}[V(ID_{Sensor}[Value])\\
\nonumber=P(ID_{Sensor}[Value])P(ID_{Sensor}[Value])
\end{eqnarray}}
Since $w_{value} = 1$, according to Eq. (10).
As for probability $P(Ob_{Sensor})$, according to Eq. (14), we have:

\begin{eqnarray}
P(Ob_{Sensor})=P(Tag_{Ob})=P(ID_{M*N})=\\
\nonumber \frac{P(ID_{1N}+...+P(ID_{MN}))}{M}
\end{eqnarray}

Using Eqs.(12) and (10) we can compute probability $P(ID_{1N})$ as:
\begin{equation}
P(ID_{1N})=1-\prod_{i=1}^{n}(1-P(ID_{1i}[Value]))
\end{equation}
A similar equation can be obtained from $P_(ID_{2N})$ to $P(ID_{MN})$.
Finally, join Eqs. (16) through (19), we have:
{\tiny
\begin{eqnarray}
P(Sensor)=P(ID_{Sensor})P(Ob_{Sensors})=P(ID_{Sensor}[Value])\times \\
\nonumber (\frac{1-\prod_{i=1}^{n}(1-P(ID_{1i}[Value]))+ ...+1-\prod_{i=1}^{n}(1-P(ID_{mi}[Value]))}{M}
\end{eqnarray}}

\section{Simulation}
We measure the PML similarity in terms of timing results and effectiveness on data, which is followed XML Schemas of PmlCore.xsd and Identifier.xsd and is generated randomly by PML generator. Our evaluation covers (1) timing result for various sizes of PML documents, (2) the impact of various sizes of PML documents on effectiveness, and (3) the impact of various sizes of the same elements in PML documents on effectiveness.
\subsection{Data Sets}
We use four different data sets.
\begin{table}[htbp]
	\caption{Data Sets}
   \setlength{\abovecaptionskip}{0pt}
\setlength{\belowcaptionskip}{10pt}
	\label{table 6}
	\begin{center}
	\begin{tabular}{|c|c|c|} \hline
		Num.Data Set & Degree of duplicates \\ \hline
	Data Set $1$ & $500$ random PML documents \\ \hline
     Data Set $2$ & $20$\% of the same 500 PML documents \\ \hline
    Data Set $3$ & $50$\% of the same 500 PML documents \\ \hline
    Data Set $4$ & $80$\% of the same 500 PML documents\\ \hline
		\end{tabular}
	\end{center}
\end{table}

And theses data sets are extracted from PML data generator designed by our project team, which enable to generate different PML documents in accordance with our needs.
In the generator, the parameters of self-definition include (1) amount of Tag element, (2) type of Tag element, and (3) value of ID element. Hence, Dataset 1 represents the scenario where we don’t understand the structure and duplicate of PML documents, and all of theses PML documents are randomly generated. Dataset 2, 3, 4 are used to show the impact of different degree of duplicates to timing result and effectiveness of our algorithm.

These tests were done on a Thinkpad X220i computer with dual processor CPU of Core i3 2370M Processors, running at 2.4 GHz. All simulative approaches, include measure of timing result and effectiveness, were implemented by us in Matlab. And we know that the timing results of algorithm could be influence by different computer.
In this section we describe some simulations that measure the PML similarity in terms of timing results and effectiveness on data, which is followed XML Schemas of PmlCore.xsd and Identifier.xsd and is generated randomly by PML generator. We have three main goals in our simulations:
\begin{itemize}
\item Probability mean—the average of duplicate probability.We use the range of probability mean to evaluate what is the value of final probability that could be considered as duplicates for two PML documents.
\item Time performance—the timing result of our algorithm for different amounts of duplicate data.We use the data to evaluate the timing result of our algorithm in accordance to different scenarios.
\item Precision and recall values—the standard for evaluating information retrieval methods.We use recall/precision curve \cite{22} to evaluate the impact of various sizes of the same elements in PML documents on effectiveness
\end{itemize}
\subsection{Simulative Setup}
Firstly, we define the prior probability as follow.

\begin{eqnarray}
P(ID_{ij}[Value])=Sim(ID_{i},ID_{j})=\\
\nonumber 1-\frac{Compare(ID_{i},ID_{j})}{Max(|ID_{i}|,|ID_{i}|)}
\end{eqnarray}
Where $Compare(ID_{i},ID_{j})$ presents the comparison of strings $ID_{i}$ and $ID_{j}$ and the result is the integer value of difference of two strings. $|ID|$ is the length of string ID. So the result of $Max(|ID_{i}|,|ID_{j}|)$ is the maximum value of two strings.
To measure effectiveness, we use the commonly used precision and recall as presented by \cite{19}. Precision measures the percentage of correctly identified duplicates contained over the total set of objects determined as duplicates by the system. Recall measures the percentage of duplicates correctly identified by the system over the total set of duplicate objects as presented by \cite{20}.
\subsection{Simulations}
{\bf Simulation 1} to measure what the value of final probability could be considered duplicates for two PML documents by using Data Set 1.
Firstly, we should determine whether a distribution of statistics follows a normal distribution compared with the probability density function of normal distribution graph. The frequency histograms are constructed in Figure 7.
\begin{figure}
\centering
\includegraphics[width=9cm, height=8cm]{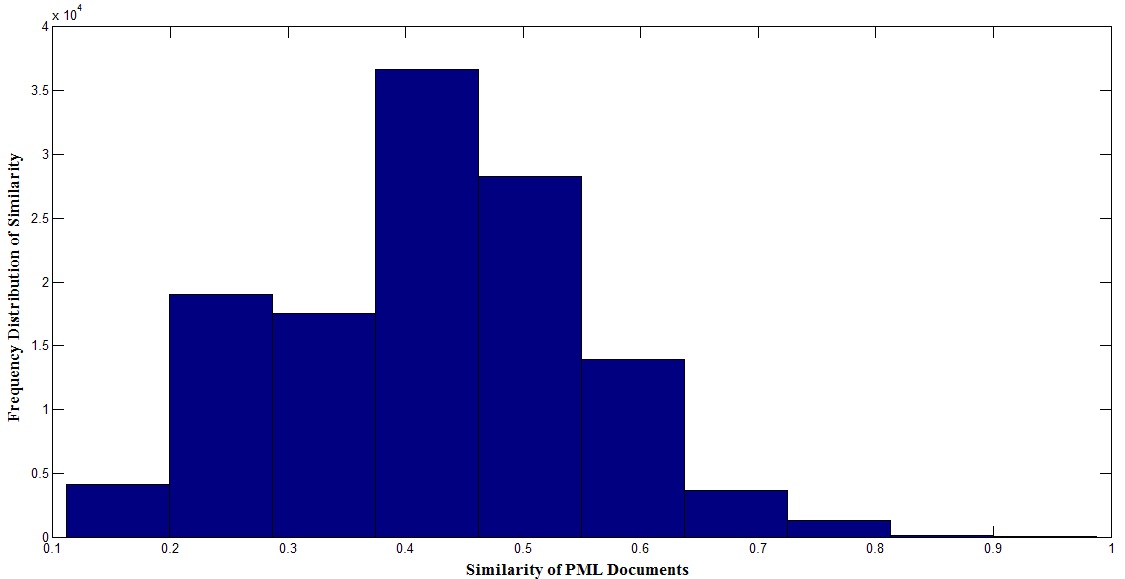}
\caption{Frequency histogram}
\label{Frequency histogram}
\end{figure}
Secondly, a normal distribution could be verified by Figure 8. With the increase of Data Set, the discrete points close to the inclined straight line segments. So the conclusion is that the values of final probability approximate normal distribution.

\begin{figure}
\centering
\includegraphics[width=9cm, height=7cm]{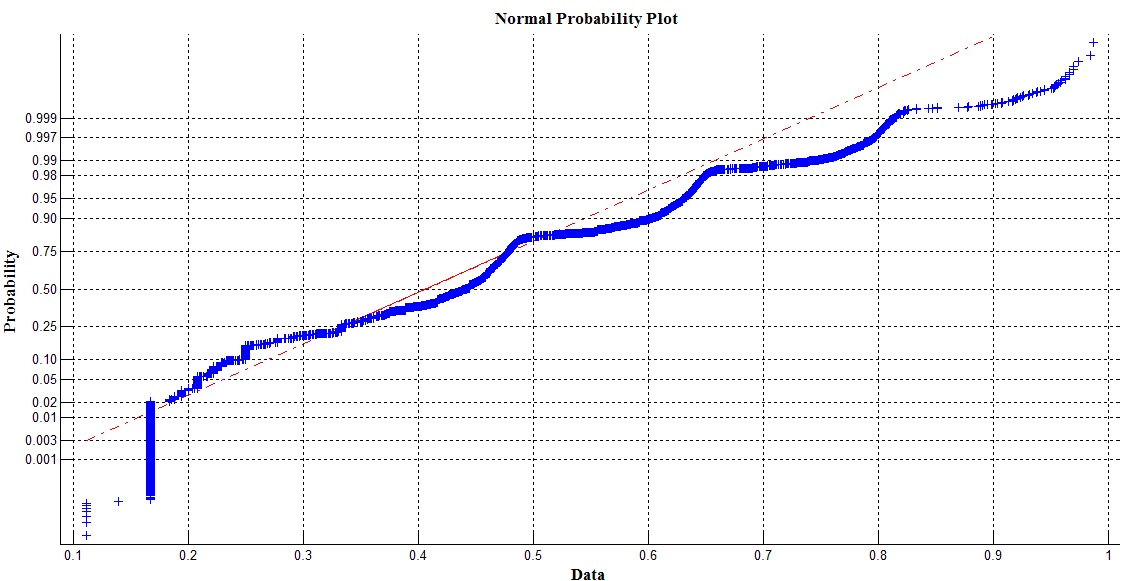}
\caption{Distribution normality test}
\label{Distribution normality test}
\end{figure}

Finally, we perform three sets of random experiments, the average of the data show as follows.
\begin{figure}
\centering
\includegraphics[width=9cm, height=7cm]{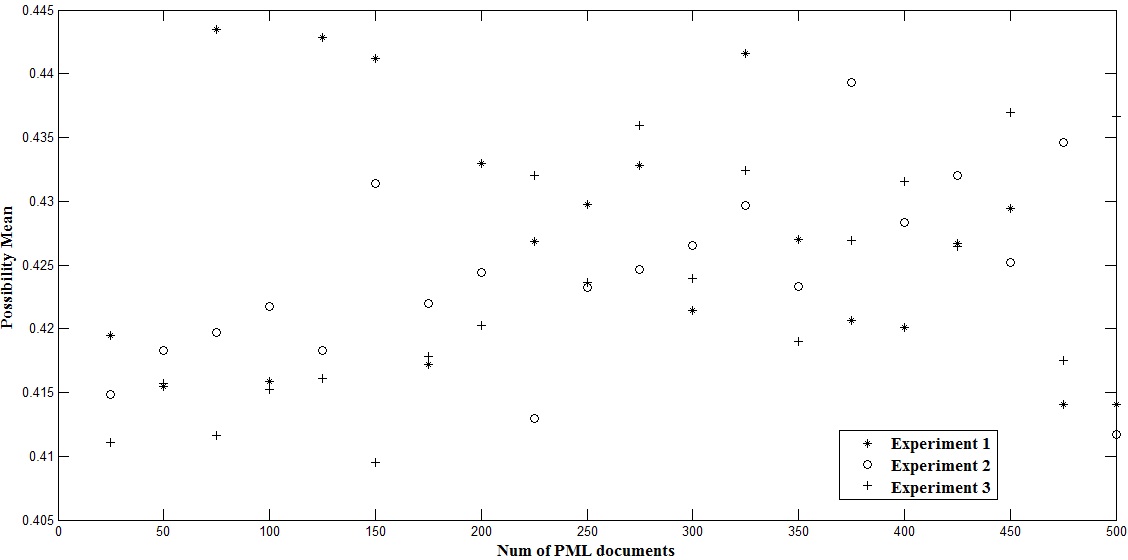}
\caption{Range of probability mean in three random simulations}
\label{Range of probability mean in three random simulations}
\end{figure}

From the Figure 9, the average mostly concentrates in the [0.4095, 0.4435]. So only objects whose duplicate probability is above or equal to the value range [0.4095, 0.4435] are considered similarity.

{\bf Simulation 2} to evaluate the timing result of our algorithm in accordance to different scenarios by using Dataset 1-4. From the Figure 10, the time to compare pairs of PML documents of various sizes grows in an almost perfect linear fashion with size and duplicate of PML documents.
\begin{figure}
\centering
\includegraphics[width=9cm, height=7cm]{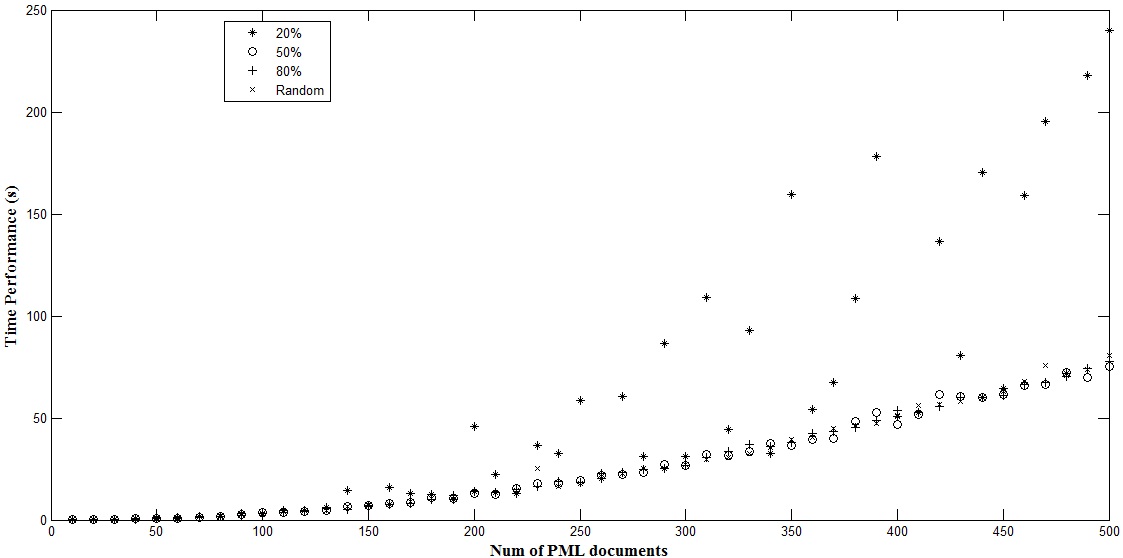}
\caption{Time performance for different amounts of duplicate data}
\label{Time performance for different amounts of duplicate data }
\end{figure}

{\bf Simulation 3} to evaluate the impact of various sizes and duplicates PML documents on effectiveness. The simulation was performed to determine the impact of the quality of the data being processed on the performance of the Bayesian network model. Figure 11 shows the results for varying the probabilities of $20\%$, $50\%$ and $80\%$ respectively.
\begin{figure}
\centering
\includegraphics[width=9cm, height=7cm]{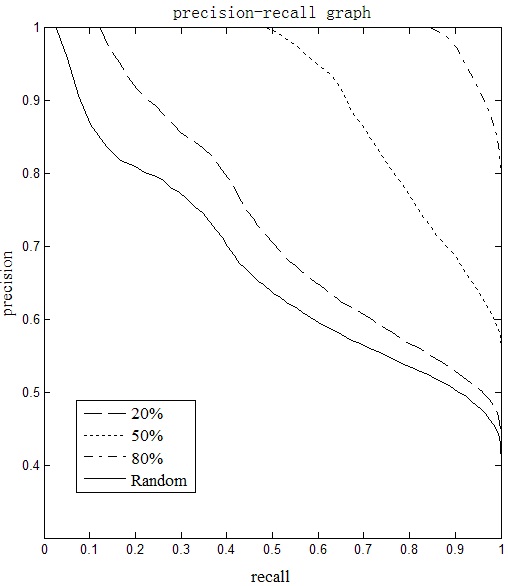}
\caption{Precision and recall values for different amounts of duplicate data}
\label{Precision and recall values for different amounts of duplicate data}
\end{figure}

\section{Conclusion}

In this paper, we introduce the function and application area of PML documents and illustrate the necessity for computing the similarity of PML documents in EPC Network. Then we propose an approach for measuring the similarity of PML documents based on Bayesian Network. With respect to the feature of PML, while measuring the similarity, we firstly reduce the redundancy data except information of EPC. On the basis of this, the Bayesian Network model derived from the structure of the PML documents being compared is constructed. And this model has accounted for both the ID values contained in the PML and their internal structure. Then the similarity between two PML documents could be deduced. Finally, the simulations evaluate the value range of similarity, timing result and the effectiveness of the similarity measure.
We intend to further validate our similarity measures by considering Real-World Data, which could exist errors, such as missing data (e.g. lack of EPC) or incompleteness data (e.g. the EPC less than 96 bit) and so on, so we still need to validate this observation.
Another issue we should intend to consider is the scalability ether in space or in time. Scaling to large amounts of PML document with the help of external memory units also needs to be studied in the future.

\section*{Acknowledgement}
The research is support by National Natural Science Foundation of P. R. China (Grant No. 61170065 and 61003039), Peak of Six Major Talent in Jiangsu Province (Grant No.2010DZXX026), Project sponsored by Jiangsu provincial research scheme of natural science for higher education institutions (Grant No.12KJB520009), Science \& Technology Innovation Fund for higher education institutions of Jiangsu Province (Grant No.CXZZ11-0405).

%%%%%%%%%%%%%%%%%%%%%%%%%%%%%%%%%%%%%%%%%%%%%%%%%%%%%%%%%%%%%%%%%%%%%%%%%%%%%%%%%%%%%

%\bibliography{ijmso}

\begin{thebibliography}{10}
\bibitem[\protect\citeauthoryear{Liang et al.}{2011}]{1}
Liang Zhou, Han-Chieh Chao.Multimedia Traffic Security Architecture for Internet of Things J. IEEE Network, 25(3):29-34,2011.

\bibitem[\protect\citeauthoryear{Floerkemeier et al.}{2003}]{2}
Floerkemeier C, Anarkat D, et al. PML core specification 1.0 J. Auto-ID Center Recommen-
dation, 2003, 15.

\bibitem[\protect\citeauthoryear{Clark et al.}{2003}]{3}
Clark S, Traub K, Council D A U C, et al. Auto-ID Savant Specification 1.02. 2003.

\bibitem[\protect\citeauthoryear{Floerkemeier et al.}{2003}]{4}
Brock D L, Milne T P, et al. The physical markup language J. Auto-ID Center White Paper MIT-AUTOID-WH-003, 2001.

\bibitem[\protect\citeauthoryear{Liang et al.}{2011}]{5}
Liang Zhou, Han-Chieh Chao,et al.Joint Forensics-Scheduling Strategy for Delay-Sensitive Multimedia Applications over Heterogeneous Networks J. IEEE JSAC,29(7):1358-1367,2011.

\bibitem[\protect\citeauthoryear{Liang et al.}{2012}]{6}
Ousmane Diallo, Joel J. P. C. Rodrigues, et al.Real-time Data Management on Wireless Sensor Network: a Survey J. Journal of Network and Computer Applications,35(3):1013–1021,2012.

\bibitem[\protect\citeauthoryear{Dalamagas et al.}{2006}]{7}
Dalamagas T, Cheng T, et al. A methodology for clustering XML documents by structure J. Information Systems,31(3):187-228, 2006.

\bibitem[\protect\citeauthoryear{Manning et al.}{2008}]{8}
Manning C. D., Raghavan P. Introduction to information retrieval M. Cambridge: Cambridge University Press, 2008.

\bibitem[\protect\citeauthoryear{Tekli et al.}{2009}]{9}
 Tekli J, Chbeir R. An overview on XML similarity: background, current trends and future directions J. Computer science review,3(3):151-173, 2009.

\bibitem[\protect\citeauthoryear{Liang et al.}{2005}]{10}
 Liang W, Yokota H. LAX: An efficient approximate XML join based on clustered leaf nodes for XML data integration J. Database: Enterprise, Skills and Innovation,:82-97,2005.

\bibitem[\protect\citeauthoryear{Kade et al.}{2008}]{11}
 A.M.Kade., Heuser C.A., Matching XML documents in highly dynamic applicationsb, in: Proceeding of the 8th ACM Symposium on Document Engineering, DocEng'08, Brazil,191-198,2008.

\bibitem[\protect\citeauthoryear{Kade et al.}{2005}]{12}
M. Weis, Naumann F., Dogmatix tracks down duplicates in XML, in: Proceedings of the ACM SIGMOD Conference, USA,431-442,2005.

\bibitem[\protect\citeauthoryear{Dorneles et al.}{2004}]{13}
Dorneles C.F.Heuser C.A., Lima A.E.N., Da Silva A.S., De Moura E.S., Measuring similarity between collections of values, in: Proceedings of the ACM international Workshop on Web Information and Data Management, USA,56-63,2004.

\bibitem[\protect\citeauthoryear{Dorneles et al.}{2007}]{14}
 Leitao L., Calado P., Weis M. Structure-based inference of XML similarity for fuzzy duplicate detection, in: Proceedings of the 16th ACM Conference on Information and Knowledge Management, CIKM'07, Portugal,293-302,2007.

\bibitem[\protect\citeauthoryear{Leong et al.}{2004}]{15}
Leong K. S., Ng M L, Engels D W. EPC network architecture. Auto-ID labs research workshop. Zurich, Switzerland, 2004.

\bibitem[\protect\citeauthoryear{WWW Consortium}{2009}]{16}
WWW Consortium, The Document Object Model, http://www.w3.org/DOM.

\bibitem[\protect\citeauthoryear{Zhang and Li}{2003}]{17}
Zhang Z.,  Li R., Similarity metric in XML documents, in: Knowledge Management and Experience Management Workshop, Germany, 2003.

\bibitem[\protect\citeauthoryear{Nierman and Jagadish}{2002}]{18}
Nierman A., Jagadish H.V., Evaluating structural similarity in XML documents, in: Proceedings of the 5th ACM SIGMOD International Workshop on the Web and Databases, WebDB,61-66,2002.

\bibitem[\protect\citeauthoryear{Pearl}{1998}]{19}
Pearl J. Probabilistic Reasoning in Intelligent Systems: Networks of Plausble Inference M. Morgan Kaufmann Pub, 1988.

\bibitem[\protect\citeauthoryear{Silander and Myllymaki}{2012}]{20}
Silander T, Myllymaki P. A simple approach for finding the globally optimal Bayesian network structure J. arXiv preprint arXiv:1206.6875, 2012.

\bibitem[\protect\citeauthoryear{Ben Gal}{2007}]{21}
 Ben Gal I. Bayesian networks J. Encyclopedia of statistics in quality and reliability, 2007.

\bibitem[\protect\citeauthoryear{Davis and Goadrich}{2006}]{22}
Davis J, Goadrich M. The relationship between Precision-Recall and ROC curves. Proceedings of the 23rd international conference on Machine learning. ACM,233-240,2006.

\bibitem[\protect\citeauthoryear{Manning et al.}{2008}]{23}
 Manning C D, Raghavan P, Sch�tze H. Introduction to information retrieval M. Cambridge: Cambridge University Press, 2008.


 \end{thebibliography}
%\bibliographystyle{unsrt}
%\bibliographystyle{alpha}

\end{document}